# Atomic mercury vapor inside a hollow-core photonic crystal fiber

Ulrich Vogl,[1,2,*] Christian Peuntinger,[1,2] Nicolas Y. Joly,[2,1] Philip St.J. Russell,[1] Christoph Marquardt,[1,2] and Gerd Leuchs[1,2]

[1]*Max Planck Institute for the Science of Light, Günther-Scharowsky-Str. 1/Bldg. 24, 91058 Erlangen, Germany*
[2]*Institute for Optics, Information and Photonics, Universität Erlangen-Nürnberg, Staudtstraße 7/B2, 91058 Erlangen, Germany*
*\*Corresponding author: ulrich.vogl@mpl.mpg.de*

**Abstract:** We demonstrate high atomic mercury vapor pressure in a kagomé-style hollow-core photonic crystal fiber at room temperature. After a few days of exposure to mercury vapor the fiber is homogeneously filled and the optical depth achieved remains constant. With incoherent optical pumping from the ground state we achieve an optical depth of 114 at the $6^3P_2$ - $6^3D_3$ transition, corresponding to an atomic mercury number density of $6 \times 10^{10}$ cm$^{-3}$. The use of mercury vapor in quasi one-dimensional confinement may be advantageous compared to chemically more active alkali vapor, while offering strong optical nonlinearities in the ultraviolet region of the optical spectrum.

## References and links

1. P. St.J. Russell, "Photonic-crystal fibers," J. Lightwave Tech. **24**, 4729–4749 (2006).
2. A. Furusawa and P. Van Loock, Quantum Teleportation and Entanglement: a Hybrid Approach to Optical Quantum Information Processing (John Wiley & Sons, 2011).
3. D. G. Angelakis, M. X. Huo, D. Chang, L. C. Kwek, and V. Korepin, "Mimicking Interacting Relativistic Theories with Stationary Pulses of Light," Phys. Rev. Lett. **110**, (2013).
4. D. E. Chang, V. Gritsev, G. Morigi, V. Vuletic, M. D. Lukin, and E. A. Demler, "Crystallization of strongly interacting photons in a nonlinear optical fibre," Nat. Phys. **4**, 884–889 (2008).
5. O. Firstenberg, T. Peyronel, Q. Y. Liang, A. V. Gorshkov, M. D. Lukin, and V. Vuletic, "Attractive photons in a quantum nonlinear medium," Nature **502**, 71 (2013).
6. P. St.J. Russell, P. Hölzer, W. Chang, A. Abdolvand, and J. C. Travers, "Hollow-core photonic crystal fibres for gas-based nonlinear optics," Nat. Phot. **8**, 278 (2014).
7. V. Venkataraman, P. Londero, A. R. Bhagwat, A. D. Slepkov, and A. L. Gaeta, "All-optical modulation of four-wave mixing in an Rb-filled photonic bandgap fiber," Opt. Lett. **35**, 2287–2289 (2010).
8. M. Bajcsy, S. Hofferberth, V. Balic, T. Peyronel, M. Hafezi, A. S. Zibrov, V. Vuletic, and M. D. Lukin, "Efficient all-optical switching using slow light within a hollow fiber," Phys. Rev. Lett. **102**, 203902 (2009).
9. M. R. Sprague, P. S. Michelberger, T. F. M. Champion, D. G. England, J. Nunn, X. M. Jin, W. S. Kolthammer, A. Abdolvand, P. St.J. Russell, and I. A. Walmsley, "Broadband single-photon-level memory in a hollow-core photonic crystal fibre," Nat. Phot. **8**, 287–291 (2014).
10. A. D. Slepkov, A. R. Bhagwat, V. Venkataraman, P. Londero, and A. L. Gaeta, "Generation of large alkali vapor densities inside bare hollow-core photonic band-gap fibers," Opt. Express **16**, 18976–18983 (2008).
11. M. J. Renn, D. Montgomery, O. Vdovin, D. Z. Andersen, C. E. Wieman, and E. A. Cornell, "Laser-guided atoms in hollow-core optical fiber," Phys. Rev. Lett. **75**, 3253–3256 (1995).
12. K. Dholakia. "Atom hosepipes," Contemporary Physics **39**, 351-369 (1998).
13. J. A. Pechkis, and F. K. Fatemi. "Cold atom guidance in a capillary using blue-detuned, hollow optical modes," Opt. Express **20**, 13409 (2012).

14. F. Blatt, T. Halfmann, and T. Peters, "One-dimensional ultracold medium of extreme optical depth," Opt. Lett. **39**, 446–449 (2014).
15. G. Epple, K. S. Kleinbach, T. G. Euser, N. Y. Joly, T. Pfau, P. St.J. Russell, and R. Löw, "Rydberg atoms in hollow-core photonic crystal fibres," Nat. Commun. **5**:4132 doi: 10.1038/ncomms5132 (2014).
16. F. Gebert, M. Frosz, T. Weiss, Y. Wan, A. Ermolov, N. Joly, P. Schmidt, and P. Russell. "Damage-free single-mode transmission of deep-UV light in hollow-core PCF," Opt. Express **22**, 15388 (2014).
17. M. H. Keirns and S. D. Colson. "Analysis of the hyperfine structure of the mercury $6^3D_3$-$6^3P_2$ transition," Journal of the Optical Society of America **65**, 1413 (1975).
18. M. L. Huber, A. Laesecke and D. G. Friend. "Correlation for the vapor pressure of mercury," Industrial & Engineering Chemistry Research **45**, 7351 (2006).
19. E. G. Thaler, R. H. Wilson, D.A. Doughty and W. W. Beersb. "Measurement of mercury bound in the glass envelope during operation of fluorescent lamps," Journal of the Electrochemical Society, **142**(6) (1995).
20. L. Yi, S. Mejri, J. McFerran, Y. Le Coq and S. Bize. "Optical Lattice Trapping of $^{199}$Hg and Determination of the Magic Wavelength for the Ultraviolet $^1S_0$ - $^3P_0$ Clock Transition," Phys. Rev. Lett. **106**, 073005 (2011).
21. P. Villwock, S. Siol and T. Walther. "Magneto-optical trapping of neutral mercury," European Physical Journal D **65**, 251 (2011).
22. D. Kolbe, M. Scheid and J. Walz. "Triple Resonant Four-Wave Mixing Boosts the Yield of Continuous Coherent Vacuum Ultraviolet Generation," Phys. Rev. Lett. **109**, 063901 (2012).
23. S. Spälter, P. van Loock, A. Sizmann and G. Leuchs. "Quantum non-demolition measurements with optical solitons," Applied Physics B: Lasers and Optics **64**, 213 (1996).
24. W. Zhong, C. Marquardt, G. Leuchs, U. L. Andersen, P. Light, F. Couny and F. Benabid. "Squeezing by self-induced transparency in Rb filled hollow core fibers," CLEOE-IQEC 2007, doi: 10.1109/CLEOE-IQEC.2007.4386698.
25. R. Marskar and U. L. Österberg. "Backpropagation and decay of self-induced-transparency pulses," Phys. Rev. A **89**, 023828 (2014).

---

**1. Introduction**

Strong nonlinear optical effects can be obtained at low light levels by confining the optical field tightly and using media with very large optical nonlinearities. This may be achieved by using the nonlinearity of strong atomic transitions combined with tight transverse confinement of light in hollow-core photonic crystal fiber (HC-PCF) [1]. In such a set-up, both atoms and light are confined to a core ~20 μm in diameter over essentially an unlimited Rayleigh length. The use of hollow-core PCF filled with Hg vapor may provide significant nonlinearities at the single photon level, permitting for example single photon all-optical switches and near-deterministic phase-gates for single photons with phase-shifts of order π [2]. Additionally, strong lateral confinement may allow the study of the dynamics of quasi one-dimensional photon-gas systems [3,4] and interactions of single photon wave packets [5].

HC-PCF filled with noble or Raman-active gases has been used to enhance many different nonlinear effects [6], leading to a new generation of pulse compressors, bright ultraviolet light sources and systems for generating Raman frequency combs. When filled with alkali vapors, HC-PCF has been used in experiments on electromagnetically induced transparency, all-optical switching at low light levels [7,8] and in broad-band single-photon memories [9]. Previous experiments with alkali-vapor-filled HC-PCF have used primarily rubidium and cesium, which offer high vapor pressure at room temperature and are standard species used in laser cooling. Alkali vapors have however a not-well-understood tendency to bind with the glass, which makes it difficult to load the fibers and to maintain a permanent high vapor pressure inside the core. Reasonable vapor pressures and high optical depths have been achieved for short times using light-induced atomic desorption in a warm vapor [10], or by filling the core with cold atoms extracted from a magneto-optical trap [11-14]. Very recently a HC-PCF filled with cesium vapor has been used to form Rydberg atoms [15].

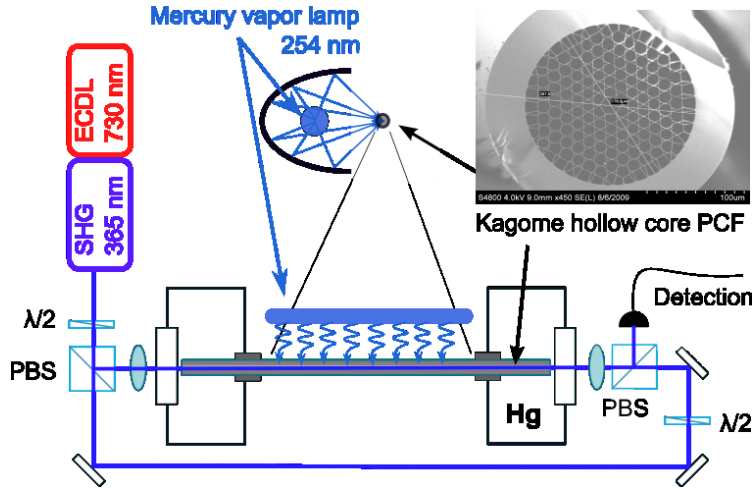

Fig. 1. Experimental set-up. The mercury vapor in the hollow-core PCF is pumped incoherently from the side by a mercury vapor lamp (main wavelength 254 nm, with significant contributions at 405 nm, 436 nm, and 546 nm), which populates the $6^3P_2$ level. The $6^3P_2$ - $6^3D_3$ transition is probed with frequency-doubled light from a diode laser.

## 2. Experimental Scheme

In this work we demonstrate high mercury vapor pressure inside a kagomé-style HC-PCF (kagomé-PCF) with a core diameter of 18 μm. Kagomé-PCFs guide light by anti-resonant reflection at the core walls [6] and offer a transmission window several hundred nm wide with losses of order 1 dB/m, even in the UV [16]. The fiber was drawn by the standard stack-and-draw technique from high-purity commercial silica tubes [1]. This multistep technique allows good homogeneity over very long length. Note that no particular treatment is done here to modify or enhance the surface roughness of either the used tubes or at any further stage of the fabrication process. A sketch of the experimental set-up is shown in Fig. 1.

A scanning electron microscope image of the kagomé-PCF used is shown in the inset of Fig. 1. The ends of a 15 cm piece of fiber were mounted inside high vacuum chambers with optical access through flanged silica windows. A small drop of mercury (natural isotope mixture) was placed in one chamber, which was then evacuated down to ~$10^{-6}$ mbar. The system was kept at room temperature (21°C).

A 5 cm length of the fiber lay bare between the two vacuum chambers and was used to pump the mercury atoms out of the ground state via the $5^1S_0$ - $6^3P_1$ transition (see Fig. 2). This was achieved with incoherent 254 nm light from a low pressure mercury vapor lamp (Deconta 25/505-21L), which was focused on to the side of the fiber with an elliptical reflector. The intensity of the 254 nm light from the vapor lamp at the fiber was ~50 mW/cm$^2$, which is similar to the saturation intensity (10 mW/cm$^2$) of the $5^1S_0$ - $6^3P_1$ transition. The fiber was uniformly illuminated over its length.

The vapor lamp additionally emitted light at the 436 nm transition ($6^3P_1$ - $7^3S_1$), which caused the long-lived $6^3P_2$ level to be populated (see Fig. 2). This was used to probe the absorption of the $6^3P_2$ - $6^3D_3$ transition with frequency doubled light from a diode laser system. The system yielded ~1 mW of linearly polarized light at 365 nm wavelength. The light was coupled into the PCF, and the input power attenuated below the saturation intensity of 55 mW/cm² for the subsequent measurements (~30 nW before the input, ~5 nW after the output, off-resonance, including all losses).

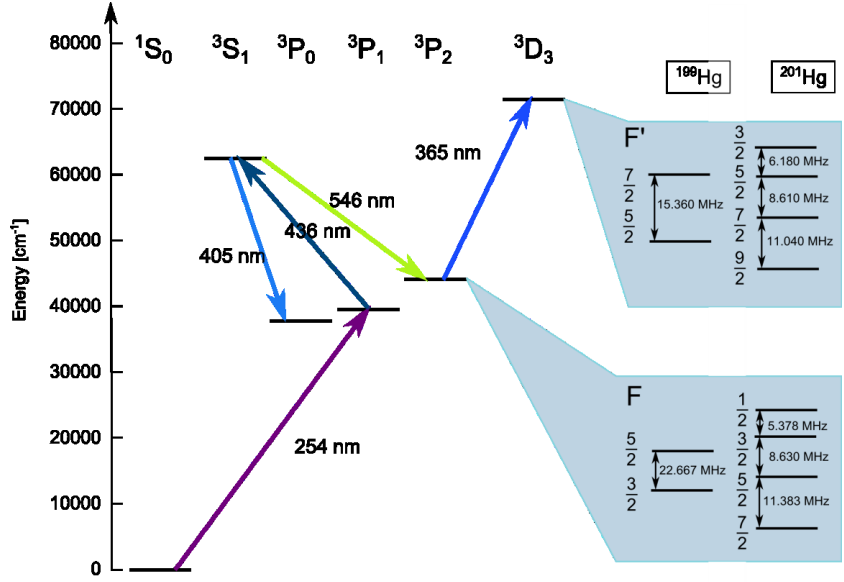

Fig. 2. Grotrian diagram of the relevant atomic mercury levels and transitions and the hyperfine splitting of the fermionic isotopes of the transition probed in the experiments. The 365 nm transition is probed with a laser, while the other depicted transitions are driven incoherently with light from a mercury vapor lamp.

## 3. Results

A scan over the Doppler-broadened $6^3P_2$ - $6^3D_3$ transition is shown in Fig. 3. It shows several transitions due to the hyperfine splitting of the fermionic isotopes $^{199}$Hg and $^{201}$Hg and contributions from the bosonic isotopes $^{198}$Hg, $^{200}$Hg, $^{202}$Hg, and $^{204}$Hg. With the known oscillator strengths for the various transitions, the relative abundance of the isotopes [16] and Doppler broadening at room temperature, we can model the transmission line shape: $I_{trans}(f) = I_0 \exp\left(-d \times \sum HF_i \times V\left[f - f_i; \gamma_D, \gamma\right]\right)$, where $d$ is the effective absorption coefficient, $V\left[f - f_i; \gamma_D, \gamma\right]$ denotes the Voigt profile of the individual hyperfine transitions due to the combined homogeneous broadening $\gamma$ and Doppler broadening $\gamma_D$, and $HF_i$ denotes the strengths of the individual hyperfine transitions [17]. In Fig. 3 we show the experimental data (red) together with the model (green); they are in excellent agreement. The temperature was treated as free parameter in the model, the best fit being at 22°C, consistent with the ambient room temperature. A maximum optical depth of 114(8) was reached at the transition of the $^{202}$Hg isotope. Using the ratio of Doppler and natural linewidth $\gamma_D / \gamma \sim 5.5$ we infer an atomic number density of $6 \times 10^{10}$ cm$^{-3}$ in the $6^3P_2$ level, the total number of Hg atoms optically probed being ~2.4×10⁶. This corresponds to about 0.1% of the saturated vapor pressure at room temperature of ~4 × 10¹³ cm⁻³ [18]. We attribute the disparity to incomplete pumping in the present scheme into the $6^3P_2$ level,

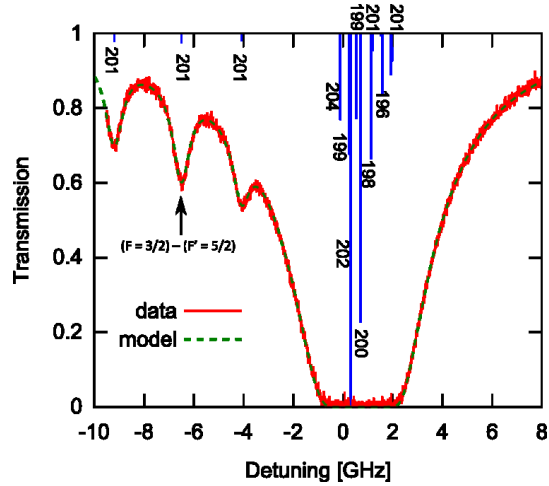

Fig. 3. Transmission spectrum through the PCF at the $6^3P_2$ - $6^3D_3$ transition at 365 nm (red). The green line is the Doppler-broadened line profile based on the known position and strength of the hyperfine transitions of the involved isotopes $^{198}$Hg, $^{199}$Hg, $^{200}$Hg, $^{201}$Hg, $^{202}$Hg, and $^{204}$Hg (blue bars).

which involves two transitions driven by the incoherent light provided by the mercury vapor lamp.

In experiments with alkali vapors the vapor density along the fiber is usually non-uniform, most of the atoms residing near the fiber ends due to slow diffusive loading and adsorption on the inner core surface. In these experiments one has also to be careful when stating an optical depth, as atoms in the cell in front of the fiber ends may also significantly contribute to the observed optical depth. A gradient in the number density also complicates the realization and description of schemes relying strongly on the local nonlinearity. The fact that we prepared the population in the $6^3P_2$ level via incoherent pumping along the side of the PCF allows us to probe directly how uniform the atomic number density is along the fiber length. If the atoms are assumed to be distributed uniformly along the fiber length L, we expect the transmitted power to directly scale with the fiber length according to the Lambert-Beer law: $I_{trans}(L) = I_0 \exp(-L_{eff}\rho) = I_0 e^{-OD}$ where $L_{eff}$ is the effective transition cross-section (accounting for Doppler broadening) and $\rho$ the atomic number density.

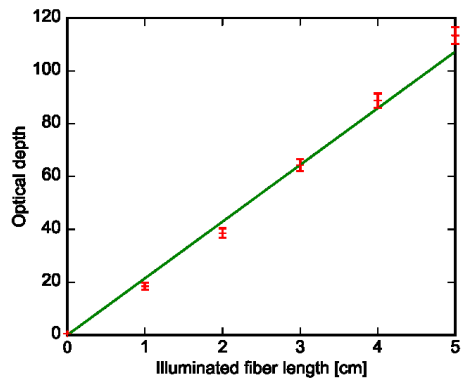

Fig. 4. Maximum optical depth of the mercury filled PCF versus the length of the fiber where mercury is pumped into the $6^3P_2$ level.

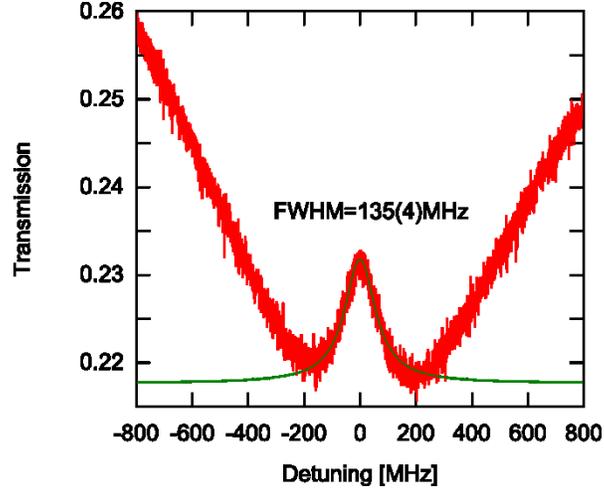

Fig. 5. Saturation spectroscopy on the (F = 3/2) – (F' = 5/2) transition of $^{201}$Hg.

By partially covering the exposed fiber length, it is possible to vary the effective length of fiber over which the $6^3P_2$ level is populated by illumination from the mercury vapor lamp. The observed optical depth versus illuminated fiber length is in good agreement with the Lambert-Beer law, as can be seen from Fig. 4, indicating a uniform atomic number density along the whole fiber length. This result shows that mercury vapor had filled the entire fiber uniformly within a few days. Furthermore, the optical depth was found to be constant over the span of the experiment (several hours). This may be due either to the side-illumination technique (which may provide constant light-induced desorption) or to the mercury atoms being generally less prone to stick to the fiber surface. In either case, the observed constant vapor pressure is an advantage compared to the alkali vapor systems previously reported.

In Fig. 5 we show a Doppler-free scan resulting from saturation absorption spectroscopy of the (F = 3/2) - (F' = 5/2) transition of $^{201}$Hg. For this, additional to the weak probe beam, a counter-propagating pump beam (~300 nW) was coupled into the fiber, derived from the same laser, and both were scanned over the Doppler-broadened transition. The Doppler-free dip was fitted with a Lorentzian function with a full-width-half-maximum of 135(4) MHz, which is in agreement with the natural linewidth of the transition of 133 MHz. Within the uncertainty of the measurement, there was no apparent transit-time broadening of the mercury atoms inside the fiber.

## 4. Conclusion

In conclusion, hollow-core photonic crystal fiber can be diffusively filled with atomic mercury vapor at room temperature, equilibrium being reached after a few days of exposure to a constant vapor pressure. An optical depth of 114(8) on the $6^3P_2 - 6^3D_3$ transition was measured, the value being limited mainly by incomplete pumping into the probed state. A much higher optical depth could be reached if the $5^1S_0$ state is probed with 253 nm light, when all mercury atoms would automatically be initially in the ground state, or simply by using a longer fiber.

The kagomé-PCF system represents a versatile room-temperature fiber-cell with high vapor pressure. By splicing and sealing the ends of the mercury-filled PCF to conventional fibers, the system could be just as easy to use as a standard all-solid glass optical fiber. Long-term investigations on the mercury-glass interactions in fluorescent lamps indicate that significant binding of mercury occurs primarily when operating with ignited plasma in the lamp, while shelved mercury vapor lamps show only very slow diffusion of mercury into the bulk glass [19].

Mercury can be laser cooled and has a long-lived clock transition at 265.6 nm [20, 21], which may make Hg-filled fibers a useful and robust tool for frequency metrology and a good platform for small and stable secondary optical standards. Nonlinear interactions might be applied to produce a tunable narrow-band UV-source, e.g., via collinear four-wave mixing or by using Hg vapor as an active medium for lasing [18, 22].

The quasi-1D confinement of the optical mode in kagomé-PCF can be used to investigate strongly interacting photonic quantum fluids. The potentially high optical depth and single-photon cooperativity achievable in this system, combined with an optimized PCF structure, may allow observation of effects such as photon-crystallization or the optical quantum-simulation of Hamiltonians relevant in solid state physics [4]. The transition investigated here (for the bosonic isotopes) is a reasonable approximation to a two-level system, as is the even stronger $6^1S_0$ - $6^3P_1$ transition directly from the ground state, which may both be suitable for investigating soliton dynamics and self-induced transparency phenomena [23–25].

Acknowledgment
We thank Phil Bucksbaum for suggesting the use of Hg vapor in hollow core PCF.